\documentclass{ws-ijbc}
\usepackage{graphicx}
\usepackage{epstopdf}
\begin{document}

\catchline{}{}{}{}{} 

\markboth{R. I. Sosa $\&$ D. H. Zanette}{Multistability of Globally Coupled Duffing Oscillators}

\title{MULTISTABILITY OF GLOBALLY COUPLED \\ DUFFING OSCILLATORS}

\author{RA\'UL I. SOSA and DAMI\'AN H. ZANETTE \footnote{Also at Consejo Nacional de Investigaciones Cient\'{\i}ficas y T\'ecnicas (CONICET), Argentina.}}
\address{Centro At\'omico Bariloche and Instituto Balseiro \\
Comisi\'on Nacional de Energ\'{\i}a At\'omica and Universidad Nacional de Cuyo \\
Av. Ezequiel Bustillo 9500, 8400 San Carlos de Bariloche, R\'{\i}o Negro, Argentina \\
iansosa996@gmail.com, zanette@cab.cnea.gov.ar}

\maketitle

\begin{history}
\received{(to be inserted by publisher)}
\end{history}

\begin{abstract}
We analyze the collective dynamics of an ensemble of globally coupled, externally forced, identical mechanical oscillators with cubic nonlinearity. Focus is put on solutions where the ensemble splits into two internally synchronized clusters, as a consequence of the bistability of individual oscillators. The multiplicity of these solutions, induced by the many possible ways of distributing the oscillators between the two clusters, implies that the ensemble can exhibit multistability. As the strength of coupling grows, however, the two-cluster solutions are replaced by a state of full synchronization. By a combination of analytical and numerical techniques, we study the existence and stability of two-cluster solutions. The role of the distribution of oscillators between the clusters and the relative prevalence of the two stable solutions are disclosed.
\end{abstract}

\keywords{Duffing oscillator, global coupling, synchronization, multistability}


\section{Introduction}

\noindent Multistability --namely, the coexistence of two or more stable stationary orbits for a given parameter set in a nonlinear dynamical system-- plays a key role in modelling several natural processes, ranging from phase transitions \cite{hyst} to biological diversification \cite{bdiv}. The possibility of selecting among a set of stationary orbits by a suitable choice of the initial condition, or switching between orbits by perturbing the motion, is also functional to numerous up-to-date technological applications. These include, for instance, quantum-based computing systems \cite{qcomp} and semiconductor optical devices \cite{semic}. In an ensemble of coupled dynamical systems, multistability underlies self-organized clustering, where the ensemble spontaneously splits into internally synchronized groups, each of them with its own trajectory. These clustered states possess rich dynamical properties \cite{mf1},
reminiscent of such complex systems and processes as neural networks \cite{Abarb,Tass}, biochemical reactions \cite{Kaneko}, and cell differentiation \cite{K2}.

Because of their interest from the viewpoints of theory and applications, multistable dynamical systems keep getting attention in several areas of research. In the last few years, a major area of application of multistability notions has been life sciences. A list of specific subjects, by no means exhaustive, comprises ecosystems \cite{ecol1},
neural networks and brain function \cite{nn1,nn3}, 
perception \cite{perc1,perc3}, 
cell function and growth \cite{cell1,cell2}, and synthetic genetics \cite{gen}. A technological field of extensive current application of those same notions is optoelectronics \cite{opt0,opt2,opt3}. 
In this field, as well as in other technological applications where the existence of multiple stable states can be pernicious, the control and suppression of multistability constitutes a topic of research by itself \cite{cont1,cont3,opt2}. 
More on the side of fundamental research, multistability has recently been studied in high-dimensional flows  \cite{hd1,hd2}, 
and in generalizations of well-known dynamical systems such as the van der Pol-Duffing oscillator \cite{chud}, the Lorenz system \cite{chen0}, 
and piecewise-linear dissipative maps \cite{oth}, among others. 

The connection between bistability in a single dynamical system and multistability in an ensemble of coupled systems of the same kind was advanced for a class of mechanical oscillators with nonlinear friction and restoring force \cite{chud}, and subsequently demonstrated for Duffing oscillators \cite{multi2}. In this latter study, it was shown that a ring of three unidirectionally coupled Duffing oscillators exhibits a rich bifurcation scenario, including regions where two or more stable solutions coexist. Likely, this multistability is a direct consequence of combined nonlinearities in the individual dynamics and in the coupling function. Consideration of the Duffing oscillator is interesting not only due to its paradigmatic role in the study of nonlinear mechanical systems \cite{Drazin}, but also because of its relevance in applications to micro-technology. A favorite design for mechanical micro-oscillators, used in pacemaking devices \cite{nano3} and in sensors \cite{Oka,sens}, consists of a tiny material beam clamped at the two ends \cite{Cleland,Ekinci}. The main nonlinear response of these clamped-clamped beams to an external excitation is well described by a mechanical oscillator subject to a cubic force of the same sign as the elastic restoring force. \cite{Nara,Tufi}. Namely, they behave as a Duffing oscillator with a {\em hardening} nonlinearity, where the restoring force grows faster than linearly with the oscillation amplitude \cite{Lif,ncomm3,Polunin}. 

In this contribution, we characterize multistability in an ensemble of  identical Duffing oscillators coupled all-to-all by an elastic interaction, and excited by an external harmonic force. As it is well known, for each suitably chosen parameter set, a harmonically forced Duffing oscillator can perform two possible stationary stable oscillations \cite{Drazin,Nayfeh}. The two oscillations have the same frequency as the excitation, but differ from each other in their amplitude and phase shift with respect to the external forcing.  As a consequence of this individual bistability, an ensemble of {\em uncoupled} Duffing oscillators obviously exhibits multistability. This corresponds to the various configurations in which the oscillators can be distributed between their two accessible stationary oscillations, forming two internally synchronized clusters with mutually different orbits. The main question addressed in the present paper is whether these two-cluster stationary states persist when the oscillators become {\it mutually coupled} by an attractive force. Attractive coupling, by which individual trajectories are led to converge to each other, is in fact the most generic mechanism leading to synchronization of interacting dynamical systems \cite{sync1,sync2}. For globally coupled bistable Duffing systems prepared to oscillate along different stable orbits, we expect that a competition emerges between their individual tendency to stay in separate trajectories and the all-to-all interaction. We demonstrate that the existence and stability of the two-cluster oscillations critically depend on the strength of coupling and on the distribution of oscillators between the two groups.

In Sect.~\ref{sec2}, we first review the main features of the individual dynamics of a bistable Duffing oscillator, and then define a mechanism of global (mean field) coupling between identical oscillators of that kind. Combining analytical and numerical techniques, we show in Sect.~\ref{sec3} that the ranges of existence and stability of two-cluster solutions sensibly differ from each other for certain parameter values. Meanwhile, they are practically coincident in other zones of parameter space. As may be expected, two-cluster oscillations eventually become unstable as the strength of coupling grows. Beyond this threshold, they are replaced by a fully synchronized solution with all oscillators belonging to a single cluster. Depending on the parameters, however, the disappearance of two-cluster solutions is associated with different kinds of critical phenomena. Results are summarized and commented in Sect.~\ref{sec4}.   

\section{Globally coupled Duffing oscillators} \label{sec2}

\subsection{Individual dynamics}

The equation of motion for a harmonically forced Duffing oscillator with spatial coordinate $x(t)$ can be written as
\begin{equation}\label{emot1}
    \ddot x+ Q^{-1} \dot x + \left( 1 +\frac{4}{3}\beta  x^2\right) x = f \cos (  \Omega t ).
\end{equation}
Here, we have chosen time units in such a way that the natural oscillation frequency  in the linear unforced limit ($\beta=0$, $f=0$) equals one. The non-dimensional quantity $Q > 0$ is the quality factor, and $\beta$ weights the nonlinear restoring force. We here consider a hardening nonlinearity, $\beta>0$ (see Introduction), but the same line of analysis can be pursued for negative $\beta$. In the right-hand side, $f$ is the forcing amplitude per unit mass and $\Omega$ is the frequency of the external force. Without generality loss, both $f$ and $\Omega$ are assumed to be positive. Note that, with our choice of time units, $f$ has the same dimensions as $x$, and $\Omega$ is non-dimensional. 

The exact solution to Eq.~(\ref{emot1}) is not known, but an efficient approximate description is provided by the multiple-scale method \cite{Nayfeh}. This approximation assumes that the typical variation times for the oscillation amplitude and phase are large as compared with the period. Generically, such condition is verified when the quality factor is large, $Q\gg 1$, so that energy dissipation occurs over relatively long time scales. To the leading order in the multiple-scale approximation, the stationary solution to Eq.~(\ref{emot1}) has the form of a harmonic oscillation with the same frequency as the external force, $x(t) = A \cos (\Omega t-\phi)$. The amplitude $A$ and the phase shift $\phi$ must satisfy
\begin{equation} \label{1osc}
\begin{array}{rl}
(1-\Omega^2 +\beta A^2) A &= f \cos \phi ,  \\
Q^{-1} \Omega A &= f \sin \phi.  
\end{array}
\end{equation}
Squaring and adding up these two equations, we get a third-order polynomial equation for $A^2$. By convention, we only pay attention to real solutions with $A > 0$. The corresponding negative amplitudes are trivially obtained by changing $\phi \to \phi\pm \pi$. As it is well known  \cite{Drazin}, three real positive solutions for the amplitude can exist within a bounded interval of values of the forcing frequency $\Omega$. This occurs when the coefficients $Q$, $\beta$, and $f$ are large enough. In qualitative terms, this is equivalent to requiring a well-developed nonlinear response to a sufficiently strong external excitation, which in turn calls for relatively weak damping.  The multiple-scale method shows that, when three solutions exists for a given forcing frequency, the solutions with the largest and the smallest amplitudes correspond to stable stationary oscillations. Meanwhile, the oscillation with intermediate amplitude is unstable \cite{Nayfeh}. Under such conditions, thus, the system is bistable. This scenario is fully confirmed by numerical integration of Eq.~(\ref{emot1}).

\begin{figure}[h]
\begin{center}
\includegraphics[width=0.6\columnwidth]{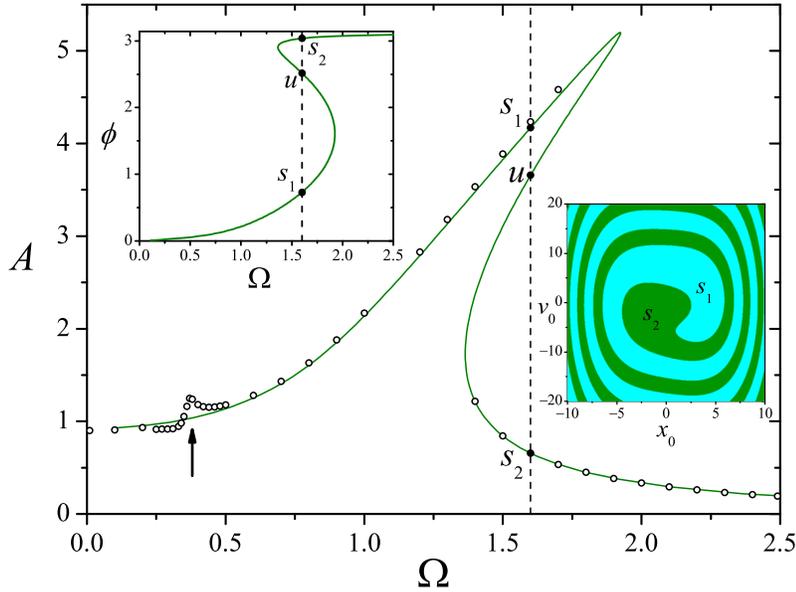}
\end{center}
\caption{Resonance curve (amplitude $A$ vs.~forcing frequency $\Omega$) of the Duffing oscillator in the harmonic approximation, Eqs.~(\ref{1osc}), for $Q=10$, $\beta=0.1$, and $f=1$. The inset to the left shows the phase shift $\phi$ as a function of the forcing frequency. Vertical dashed lines stand at $\Omega=1.6$,  for which most of the analysis of Sect.~\ref{sec3} is performed. Full dots indicate the two stable solutions, $s_1$ and $s_2$, and the unstable solution $u$ for that frequency. Open dots correspond to estimations of the amplitude, resulting from numerical integration of the equation of motion (\ref{emot1}). The arrow points to a higher-harmonic resonance. The inset to the right shows the basins of attraction of each stable oscillation, in the plane spanned by the initial coordinate $x_0$ and velocity $v_0$, for $\Omega=1.6$, as obtained from numerical results.}
\label{fig1}
\end{figure}

The main panel of Fig.~\ref{fig1} shows the resonance curve of the Duffing oscillator, namely, the oscillation amplitude $A$ as a function of the forcing frequency $\Omega$, calculated from Eqs.~(\ref{1osc}) for $Q=10$, $\beta=0.1$, and $f=1$. For these parameters, three solutions exist in an interval of forcing frequencies bounded by $\Omega_{\min}\approx 1.365$ and $\Omega_{\max} \approx 1.924$. For $\Omega=1.6$ (dashed vertical line), full dots indicate the stable solutions of maximal ($s_1$) and minimal  ($s_2$) amplitude, and the intermediate unstable solution ($u$). The inset to the left shows the phase shift $\phi$ as a function of $\Omega$, which varies between $0$ and $\pi$ as the frequency grows. Note that the solution with minimal phase ($s_1$)  is that of maximal amplitude, and {\it vice versa}.

Open dots in the main panel of Fig.~\ref{fig1} are estimations of the amplitude of stationary oscillations obtained from numerical resolution of the equation of motion (\ref{emot1}), using a standard integration algorithm (see details in Sect.~\ref{Stab}). The coincidence with the solution to Eqs.~(\ref{1osc}) is an indication of the high quality of the harmonic approximation obtained as the leading order of the multiple-scale method. Deviations only appear for large amplitudes, where nonlinear effects are expected to be more relevant, and in some specific frequency ranges, as marked by the vertical arrow just below $\Omega=0.5$. There, the small amplitude peak reveals a higher-harmonic resonance \cite{PLA} --in this case, when $\Omega$ is around $1/3$ of the oscillator's natural frequency-- which naturally escapes the harmonic approximation. In the  bistability range $(\Omega_{\min},\Omega_{\max} )$, where we focus our analysis, these higher-order effects are however absent. 

Also resulting from numerical integration of Eq.~(\ref{emot1}), in the inset to the right of Fig.~\ref{fig1} we show the basins of attraction of the two stable oscillations for $\Omega=1.6$, over the plane spanned by the initial coordinate $x_0=x(0)$ and velocity $v_0= \dot x(0)$. In Sect.~\ref{sec3}, we use this information to numerically build stationary states with a controlled number of oscillators in each  stable oscillation.

\subsection{Mean-field global coupling and two-cluster equations} \label{2clus}

A set of $N$ identical Duffing oscillators, individually satisfying Eq.~(\ref{emot1}), can be globally coupled to each other by introducing, for each oscillator $i$, a coupling force
\begin{equation}
f_i^{K} =- \frac{K}{N} \sum_{j=1}^N (x_i - x_j) =- K(x_i-\langle x \rangle ).
\end{equation}
Here, $x_i(t)$ is the coordinate of oscillator $i$ and the non-dimensional parameter $K$ is the coupling strength. This all-to-all coupling is equivalent to a linear attractive force of elastic constant $K$, centered at the average coordinate $\langle x \rangle = N^{-1} \sum_j x_j$. Consequently, its physical effect is to lead the individual coordinate of each oscillator to approach $\langle x \rangle$. The average coordinate thus plays the role of a mean field \cite{mf1,mf2}, collectively driving the oscillators towards a common trajectory. This effect, however, competes with the tendency of each oscillator to remain in its individual stable orbit. The globally coupled equations of motion read
\begin{equation}\label{emotN}
    \ddot x_i+ Q^{-1} \dot x_i + \left( 1 +\frac{4}{3}\beta  x_i^2\right) x = f \cos ( \Omega t )-K(x_i-\langle x \rangle),
\end{equation}
for $i=1,\dots , N$. 

As advanced in the Introduction, we seek stationary oscillations where the globally coupled ensemble is split into two clusters, generally, with different numbers of oscillators. While within each cluster all oscillators are mutually synchronized and follow identical orbits, the two clusters perform  oscillations with different amplitudes and phases. In the absence of coupling, this situation can be obviously attained just by distributing the ensemble between the two stable oscillations accessible to each individual element. The question thus is to which extent these collective states survive when coupling is turned on. To solve this problem to the leading order in the multiple-scale approximation, we suppose that the ensemble is divided into two clusters $1$ and $2$, respectively containing $N_1$ and $N_2$ oscillators ($N_1+N_2=N$). The clusters perform stationary harmonic oscillations with coordinates $X_{1,2} (t) = A_{1,2} \cos (\Omega t -\phi_{1,2})$. The corresponding mean field reads
\begin{equation}
    \langle x \rangle = r_1 A_1 \cos (\Omega t -\phi_1)+(1-r_1) A_2 \cos (\Omega t -\phi_2),
\end{equation}
with $r_1 = N_1/N$ ($0< r_1< N$). Under these assumptions, the multiple-scale method yields
\begin{equation} \label{Nosc}
\begin{array}{rl}
(K+1-\Omega^2 +\beta A_1^2)A_1 &= f \cos \phi_1 +K r_1 A_1 +K (1-r_1) A_2 \cos (\phi_1-\phi_2) , \\ 
(K+1-\Omega^2 +\beta A_2^2)A_2 &= f \cos \phi_2 +K (1-r_1) A_2 +K r_1 A_1 \cos (\phi_2-\phi_1) , \\ 
Q^{-1} \Omega A_1 &= f \sin \phi_1+K (1-r_1) A_2 \sin (\phi_1-\phi_2), \\  
Q^{-1} \Omega A_2 &= f \sin \phi_2+K r_1 A_1 \sin (\phi_2-\phi_1).  
\end{array}
\end{equation}
These are four algebraic equations for the amplitudes $A_1$ and $A_2$, and the phases $\phi_1$ and $\phi_2$ of the stationary oscillations performed by the two clusters. Although a full analytical treatment is not possible, the existence of solutions can be efficiently dealt with by  numerical methods. On the other hand, a full assessment of stability for the corresponding oscillations would require considering all possible perturbations from the stationary motion. In particular, it would be necessary to allow the trajectory of each individual oscillator to be perturbed. The treatment of this much higher-dimensional problem within the multiple-scale approximation is impractical, and will here be dealt with by direct numerical integration of the equations of motion   (\ref{emotN}).

\section{Two-cluster stationary oscillations} \label{sec3}

\subsection{Existence} \label{Exis}

For a given choice of parameters, and within the harmonic approximation, two-cluster stationary oscillations exist if Eqs.~(\ref{Nosc}) have real solutions for the amplitudes $A_{1,2}$ and the phases $\phi_{1,2}$. Since we must ultimately resort to numerical methods to solve the equations, our attention will be focused on the dependence on the parameters $r_1$  and $K$, which are the most relevant to the problem addressed in the Introduction. The fraction $r_1$, in fact, characterizes the multiplicity of states associated with the distribution of oscillators between the two clusters. Meanwhile,  the coupling strength $K$ is expected to control the stability of the two-cluster oscillations.  The remaining parameters are chosen in such a way that the analysis is representative of more general situations. The results presented below mostly correspond to the choice $Q=10$, $\beta=0.1$, $f=1$, and $\Omega=1.6$ (cf.~Fig.~\ref{fig1}).

As expected, in the absence of coupling ($K=0$) and for any value of $r_1$, Eqs.~(\ref{Nosc}) reduce to Eqs.~(\ref{1osc}) for each pair $A_i$, $\phi_i$ ($i=1,2$). In this limit, and within the bistability regime, the solutions for $A_{1,2}$ and $\phi_{1,2}$ are given by the nine possible combinations of the three solutions to Eqs.~(\ref{1osc}) for the individual amplitude $A$ and phase $\phi$.  These nine solutions should play the role of ``precursors'' of the solutions to Eqs.~(\ref{Nosc}) when $K$ becomes positive --at least, for small $K$-- in the sense that, when coupling is turned on, each solution should be the continuation of one of the solutions for $K=0$. Specifically, we are interested in the solutions for $A_{1,2}$ and $\phi_{1,2}$ whose precursors correspond to stable oscillations of the two clusters. To be concrete, in the following we focus the attention on the solution where, in the limit $K=0$, clusters $1$ and $2$ respectively oscillate in the stable solutions $s_1$ (maximal amplitude, minimal phase) and $s_2$ (minimal amplitude, maximal phase), as shown in Fig.~\ref{fig1}. We call this precursor $\{ s_1,s_2\}$ and denote by $S_{12}$ its continuation for $K>0$. We show below that, as $K$ grows,  solution $S_{12}$ eventually merges with another two-cluster solution and, from then on, both cease to exist.
 
\begin{figure}[ht]
\begin{center}
\includegraphics[width=0.9\columnwidth]{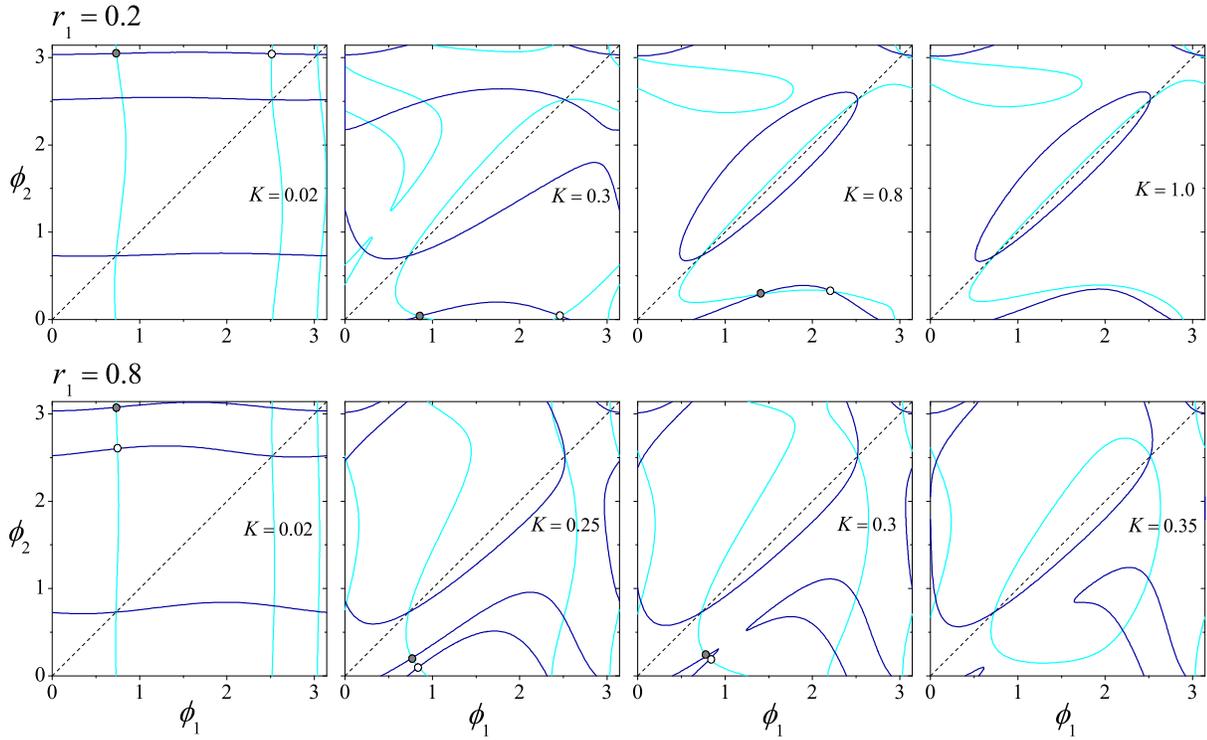}
\end{center}
\caption{The intersections of the light (cyan) and dark (blue) curves are the solutions to Eqs.~(\ref{Nosc}) for the phases $\phi_1$ and $\phi_2$. Upper and lower rows respectively correspond to $r_1=0.2$ and $0.8$, with the values of $K$ indicated inside each panel.  Other parameters are $Q=10$, $\beta=0.1$, $f=1$, and $\Omega=1.6$; cf.~Fig.~\ref{fig1}. Dotted lines indicate the diagonals $\phi_1=\phi_2$. Full and empty dots in the three first panels of each row respectively indicate the two-cluster solution  $S_{12}$, and the solution which merges with $S_{12}$ at the point where both disappear.}
\label{fig2}
\end{figure}
 
A convenient way to deal with the solutions for $A_{1,2}$ and $\phi_{1,2}$  --and, at the same time, to visualize them-- is to note that the last two lines in (\ref{Nosc}) are linear equations for $A_{1,2}$, with coefficients depending on  $\phi_{1,2}$. Therefore, the amplitudes can be immediately obtained from these two equations and replaced into the first two lines of (\ref{Nosc}), to get equations for the phases alone. These equations involve rather complicated combinations of sines and cosines of $\phi_{1,2}$ but, once a parameter set has been selected, finding their roots by the standard multivariate Newton-Raphson method is straightforward. Panels in Fig.~\ref{fig2} show the solutions for several combinations of $r_1$ and $K$, over the plane spanned by $\phi_1$ and $\phi_2$. The problem for the phases is periodic with period $\pi$. Therefore,  each axis is limited to the interval $(0,\pi)$. 

Each set of curves (plotted in light or dark shade) stands for the solutions of one of the two equations for the phases. Thus, the intersections between curves give the pairs $(\phi_1,\phi_2)$ which solve the problem. Note that, in all the panels, three intersections occur over the diagonal $\phi_1=\phi_2$. These solutions, for which we also have $A_1=A_2$,  represent the trivial cases where the two clusters have the same coordinates --i.e., where the ensemble is not split at all. Six additional intersections are seen in the panels corresponding to low values of $K$ but, as coupling grows stronger, the curves become heavily distorted and some of the intersections disappear. For sufficiently large values of $K$, in fact, we expect that the only surviving solutions are those over the diagonal, with the whole ensemble collapsed into a single fully synchronized cluster, due to the strong coupling between oscillators. 

We also point out the symmetry with respect to the diagonal between the curves of different shades in the second panel of the upper row and in the third panel of the lower row, corresponding to the same value of $K$ and complementary values of $r_1$. This symmetry reflects the invariance of the problem under the swap of labels $1$ and $2$ between the clusters, for a fixed set of parameters, which implies the exchanges $\phi_1 \leftrightarrow \phi_2$ and $r_1 \leftrightarrow 1-r_1$.  

The full dot in each one of the three leftmost panels of each row in Fig.~\ref{fig2} indicates the phases ($\phi_1,\phi_2$) corresponding to solution $S_{12}$. The empty dot, in turn, stands for the solution that, as $K$ grows, merges with $S_{12}$. In the third panel of each row the two solutions have closely approached each other, while in the fourth panel they have disappeared.  Note that the merging of the two solutions occurs though what seemingly is a typical saddle-node bifurcation scenario.   

For both values of $r_1$, the empty dot represents the continuation of a solution where, for $K=0$, one of the two clusters is in the intermediate unstable state $u$ (see Fig.~\ref{fig1}). On the one hand, for $r_1=0.2$, this solution is the continuation of $\{u,s_2\}$, in which clusters $1$ and $2$ are in states $u$ and  $s_2$, respectively. On the other, for $r_1=0.8$, it is the continuation of  $\{s_1,u \}$. Following the notation used for $S_{12}$, we call these solutions $S_{u2}$ and $S_{1u}$, respectively. The fact that, depending on the value of $r_1$, $S_{12}$ disappears by merging with a different solution points to the existence of two distinct regimes in the critical behavior of our system. In particular, the critical value of the coupling constant at which $S_{12}$ disappears, $K_D$, is sensibly different in the two cases, with $K_D\approx 0.912$ for $r_1=0.2$ and $K_D\approx 0.314$ for $r_1=0.8$. 

\begin{figure}[ht]
\begin{center}
\includegraphics[width=0.6\columnwidth]{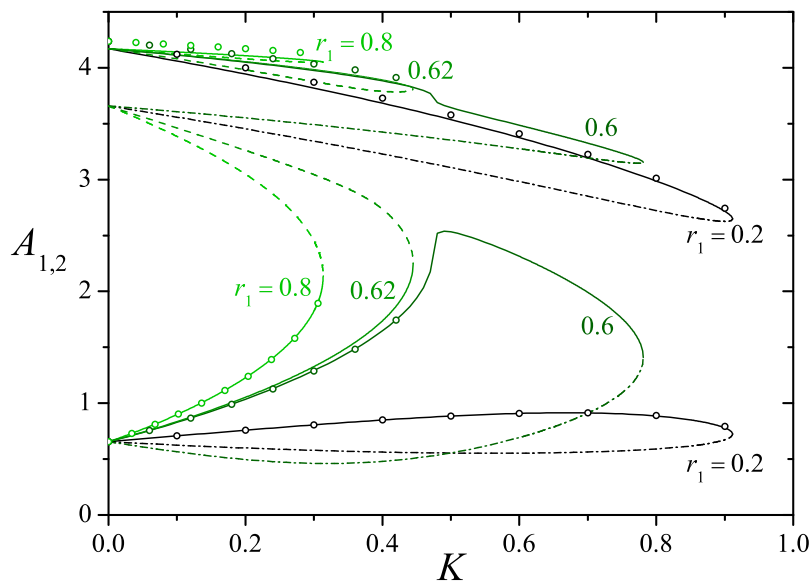}
\end{center}
\caption{Stationary oscillation amplitudes $A_{1,2}$, solutions to Eqs.~(\ref{Nosc}), as functions of the coupling strength $K$, for four values of the fraction of oscillators in cluster $1$, $r_1$. Other parameters are as in Fig.~\ref{fig2}.  Full, dashed, and dash-dotted lines respectively indicate the amplitudes corresponding to solutions $S_{12}$, $S_{1u}$, and $S_{2u}$. For large and $r_1$, the disappearance of $S_{12}$ respectively occurs by merging with $S_{1u}$ and $S_{2u}$ at the critical coupling $K_D$. Open dots stand for numerical estimations of the amplitude for $r_1=0.2$, $0.6$, and  $0.8$, obtained from  integration of the equations of motion (\ref{emotN}) for $N=100$.}
\label{fig3}
\end{figure}

The contrast between the two regimes better manifest itself in the dependence of the oscillation amplitudes on the coupling constant. Figure \ref{fig3} shows, in full lines, the amplitudes for the solution $S_{12}$ as functions of $K$, for several values of $r_1$. Other parameters are as in Fig.~\ref{fig2}. Note that, for each parameter set, the solution is represented by two curves, the upper one corresponding to $A_1$ and the lower one to $A_2$. Pairs of dashed lines with the same shade stand for the amplitudes $A_1$ and $A_2$ corresponding to the  solution $S_{1u}$. Dash-dotted lines, in turn, are the corresponding amplitudes for    $S_{u2}$. For large values of $r_1$, it is $S_{1u}$  which merges with $S_{12}$ at the critical coupling $K_D$, while for small $r_1$ merging occurs between  $S_{12}$ and $S_{u2}$. The values of $r_1$ have been chosen in such a way that two of them, $r_1=0.6$ and $0.62$, are just below and above the switching between regimes.

The amplitude $A_1$ of $S_{12}$, represented by the upper set of full curves in the figure, decreases monotonically as $K$ grows, irrespectively of the value of $r_1$, until it disappears at the corresponding critical coupling $K_D$. This behavior is consistent with the fact that, as coupling becomes stronger, the coordinates of the two clusters should approach each other. On the other hand, the amplitude $A_2$, given by the lower full curves, behaves differently depending on $r_1$. In the small-$r_1$ regime, $A_2$ grows for small $K$, reaches a maximum, and then decreases to splice itself with the branch corresponding to $S_{u2}$, coming from below. As $r_1$ increases and approaches the switching between regimes, the maximum in $A_1$ becomes higher and sharper. In the large-$r_1$ regime, in contrast, $A_1$ grows monotonically with $K$ and, at the critical coupling $K_D$, it splices itself with the dashed branch above it, corresponding to $S_{1u}$. Comparing the curves for $r_1=0.6$ and $0.62$, it becomes clear that the regime switching entails an abrupt drop of $K_D$, in the form of a finite discontinuity. 

Open dots in Fig.~\ref{fig3} are numerical estimations of the amplitudes $A_{1,2}$ of two-cluster stationary oscillations, obtained as detailed in  Sect.~\ref{Stab}, for $r_1=0.2$, $0.6$, and $0.8$ in an ensemble of $100$ oscillators. We find a good general agreement with the solutions yielded by the harmonic approximation, although --much as in Fig.~\ref{fig1}-- the discrepancy grows with the amplitude, as expected due to the effect of nonlinearity. However, for $r_1=0.6$, we note that the numerical results are limited to $K \lesssim 0.4$, while the harmonic approximation predicts that the solution reaches much larger values of the coupling constant. This substantial difference, which turns out to be related to the destabilization of $S_{12}$ in the numerical integration of Eqs.~(\ref{emotN}), is analyzed in Sect.~\ref{Stab}.

\begin{figure}[ht]
\begin{center}
\includegraphics[width=0.6\columnwidth]{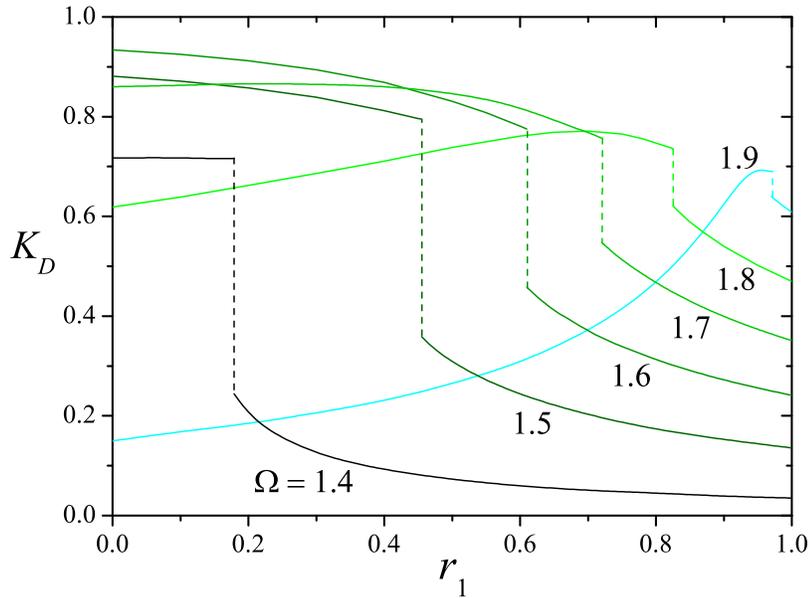}
\end{center}
\caption{Critical coupling strength $K_D$ as a function of the fraction $r_1$, for several values of the excitation frequency $\Omega$ in the bistability range. For each $\Omega$, the vertical dashed line stands at $r_1^D$, the fraction at which the change between the small- and large-$r_1$ regimes takes place. }
\label{fig4}
\end{figure}

It turns out that the overall picture described so far for $\Omega=1.6$ is also found for other values of the forcing frequency within the bistability interval $(\Omega_{\min}, \Omega_{\max})$. Naturally, however, the parameter values at which solutions disappear and regimes switch depend on $\Omega$. Figure \ref{fig4} shows the critical coupling $K_D$ as a function of $r_1$ for several values of the forcing frequency. Dashed vertical lines stand at $r_1^D$, the value of $r_1$  at which regime switching takes place. From these results, we point out the monotonous growth of $r_1^D$ as $\Omega$ increases, accompanied by a sustained decrease in the size of jump of $K_D$ at $r_1^D$. Additionally, leaving aside the effects of the discontinuity in $K_D$, we remark that the critical coupling varies non-monotonically with $\Omega$ for a given value of $r_1$. For $r_1=0.1$, for instance, $K_D$ has a relatively low value for $\Omega=1.5$, grows to attain a maximum for $\Omega\approx 1.7$, and then decreases to very low values for large $\Omega$. 

\subsection{Stability} \label{Stab}

As advanced above, we study the stability of two-cluster oscillations by directly integrating the equations of motion (\ref{emotN}) by numerical means. This is motivated by the fact that the two-cluster harmonic approximation disregards the individual degrees of freedom of  each oscillator, whose deviation from the two-cluster trajectories might destabilize those states. Indeed, the results presented in this section suggest that this is the case. On the other hand, the analytical formulation of the harmonic approximation for the full $N$-oscillator ensemble is impractical in the study of both existence and stability. 

Integration of Eqs.~(\ref{emotN}) --as well as of Eqs.~(\ref{emot1}), for the numerical results shown in Fig.~\ref{fig1}-- was performed using a standard fourth-order Runge-Kutta algorithm. The integration step was fixed at $\delta t=0.01$, namely, well below 1\% of the typical oscillation period. In the cases where we integrated Eqs.~(\ref{emotN}) for successive values of a given parameter --such as $\Omega$ for the results of Fig.~\ref{fig1}, and $K$ for those of Fig.~\ref{fig3}-- our strategy was to run the integration for each value of the parameter until the stationary oscillation was clearly established. Then, this stationary oscillation was used as initial condition for the next value. For each value of the fraction $r_1$, in turn, two-cluster initial states were prepared in the absence of coupling ($K=0$) by suitably choosing the initial coordinates of individual oscillators within each of the  basins of attraction depicted in the rightmost inset of Fig.~\ref{fig1}. Higher values of $K$ were then reached by the ``sweeping'' procedure described just above. Finally, in order to numerically test the stability of stationary solutions, we added small-amplitude random forces to the equations of motion, as a generic representation of a perturbation over all the degrees of freedom in the system. For each oscillator and at each integration step, the random force was independently drawn from a uniform distribution around zero, with a maximal absolute value of $0.01$, namely, 1\% of the amplitude of the external excitation ($f=1$). 

For the evaluation of the critical coupling strength at which two-cluster stationary oscillations become unstable, $K_U$, we run the numerical integration for each possible value of $r_1$ ($=0.01, 0.02, \dots 0.99$ for $N=100$). We started with the two-cluster state from $K=0$ and gradually increased the coupling strength by steps of size $\delta K=10^{-3}$. Keeping record of the oscillation amplitudes, we were able to detect the coupling strength at which the two-cluster solution ceased to exist. Since our system was being continually perturbed by a random force, we identified this coupling strength with $K_U$. In all cases, we observed that, for coupling strengths just above $K_U$, the two clusters collapsed into a single group. In this group, all oscillators performed fully synchronized motion (up to the minor fluctuations induced by the random force) in one of the two stable states of the Duffing equation (\ref{emot1}), $s_1$ or $s_2$ (see Fig.~\ref{fig1}). This procedure was repeated for each value of the excitation frequency $\Omega$ considered in Fig.~\ref{fig4}.

\begin{figure}[ht]
\begin{center}
\includegraphics[width=0.6\columnwidth]{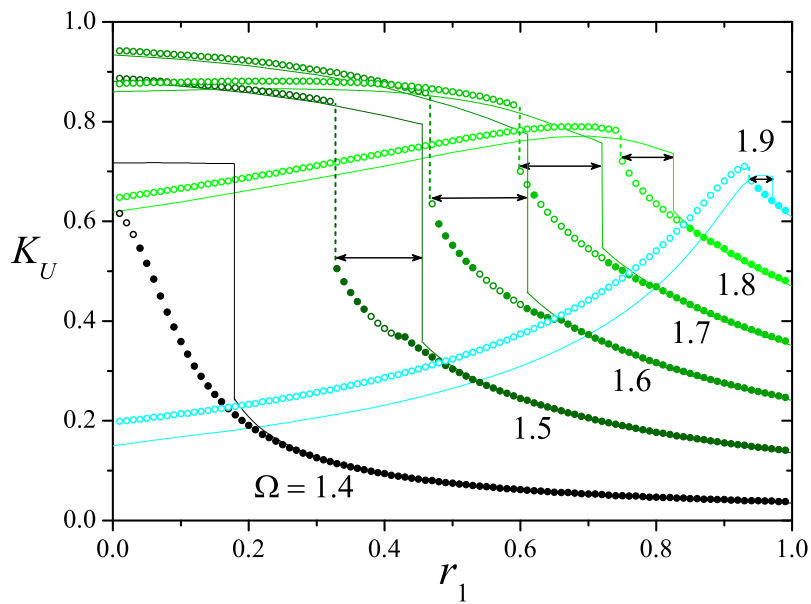}
\end{center}
\caption{Dots correspond to the numerical estimation of the two-cluster destabilization coupling strength $K_U$ as a function of $r_1$, for the same values of  $\Omega$ as in Fig.~\ref{fig4}. Full and open symbols indicates the cases where destabilization of the two-cluster solution leads the whole ensemble to states $s_1$ and $s_2$, respectively. Vertical dashed lines stand at the critical fraction $r_1^U$ where $K_U$ is discontinuous. For comparison, full thin lines show the same results for $K_D$ as in  Fig.~\ref{fig4}. Horizontal double arrows indicate the difference between $r_1^U$ and $r_1^D$, for each $\Omega\ge 1.5$. }
\label{fig5}
\end{figure}

Figure \ref{fig5} shows, as dots, the numerical estimation of the two-cluster destabilization coupling strength $K_U$ as a function of $r_1$, for each value of $\Omega$. For comparison, full thin lines show the same results as in Fig.~\ref{fig4}, namely, the critical coupling strengths $K_D$ at which two-cluster solutions cease to exist as $K$ grows. Except for $\Omega =1.4$, we find that $K_U$ has the same overall dependence on $r_1$ as $K_D$. In particular, as $r_1$ increases, $K_U$ displays a sharp drop at a critical value $r_1^U$, indicated by vertical dashed lines. This drop, however, is systematically to the left of the drop in $K_D$, i.e.~$r_1^U<r_1^D$. The horizontal double arrows mark the difference for each $\Omega \ge 1.5$. Note that if, for a given $\Omega$, $r_1$ lies in the interval $(r_1^U,r_1^D)$, there generally is a sizable difference between the coupling constant at which two-cluster solutions become unstable and the value of $K$ at which they disappear. For $\Omega=1.4$, where there is no positive value for $r_1^U$, $K_U$ is well below $K_D$ for all $r_1<r_1^D$. The large difference between $K_U$ and $K_D$ for $r_1$ in $(r_1^U,r_1^D)$ suggests that, at least within this interval, destabilization of the two-cluster solution is controlled by the perturbation of degrees of freedom other than those involved in the harmonic approximation of Sect.~\ref{2clus}.    

For $r_1$ outside the interval $(r_1^U,r_1^D)$, there is a reasonably good agreement between $K_U$ and $K_D$. This is particularly true for $r_1 > r_1^D$. For $r_1<r_1^U$, on the other hand, systematic differences appear, with $K_U$ becoming increasingly larger than $K_D$ as $\Omega$ grows. Note that, since the interval of existence of two-cluster solutions should always include that of stability, the critical coupling of destabilization should never lie above that of their disappearance. The discrepancy between $K_U$ and $K_D$ in Fig.~\ref{fig5} should therefore be ascribed to the fact that the former were obtained from direct numerical integration of the equations of motion, while the later originate in the two-cluster harmonic approximation.

Full and open dots in Fig.~\ref{fig5} indicate the cases in which, upon destabilization, the two clusters collapse to the solutions $s_1$ and $s_2$, respectively (see Fig.~\ref{fig1}). Generally, for large values of $r_1$, destabilization leads the whole ensemble to solution $s_1$ (full dots), while for small $r_1$ the collapse occurs towards $s_2$ (empty dots). For intermediate values of $r_1$, full and open dots alternate with each other in a non-systematic manner. This indicates that, in that zone, the state reached upon collapse of the two clusters sensibly depends on the initial conditions.  

\subsection{Physical interpretation}

Some of the main results obtained in Sects.~\ref{Exis} and \ref{Stab} can be semi-quantitatively understood in terms of a few arguments of physical inspiration.  These invoke, first, the dynamical dominance of either cluster according to the number of oscillators it contains. Second, they take into account the relative prevalence of the two stable solutions of the one-oscillator Duffing equation, $s_1$ and $s_2$, depending on the distance of each of them to the unstable solution $u$.  Concretely, when the fraction $r_1$ of oscillators in cluster $1$ is large, the continuation of solution $s_1$ is very similar to $s_1$ itself. In this case, cluster $1$ moves in a trajectory close to $s_1$. Conversely, when most of the oscillators are in cluster $2$, the tendency is to keep that cluster close to $s_2$.  Moreover, the motion of the largest cluster dominates the dynamics of the mean field $\langle x \rangle$ in Eq.~(\ref{emotN}). The fraction $r_1$, therefore, controls how close from either $s_1$ and $s_2$ is the largest part of the system expected to oscillate, as well as its weight in the mean-field interaction. 

In turn, the relative stability of $s_1$ and $s_2$ is controlled by the excitation frequency $\Omega$. For $\Omega \gtrsim \Omega_{\min}$, just above the lower bound of the bistability interval (see Sect.~\ref{sec2} and Fig.~\ref{fig1}), states $s_2$ and $u$ are close to each other. Therefore, the basin of attraction of $s_2$ is expected to be much smaller than that of $s_1$. {\it Vice versa}, for $\Omega \lesssim \Omega_{\max}$, $s_2$ is relatively more stable than $s_1$. In either situation, we expect that the collective motion of the oscillator ensemble stays preferentially in the vicinity of the state with stronger stability.

The existence of the two regimes in the disappearance of the two-cluster solution (see Sect.~\ref{Exis} and  Figs.~\ref{fig2}, \ref{fig3}) can be interpreted in terms of the dominance of either cluster in the collective dynamics of the whole ensemble. For small $r_1$, the cluster in the continuation of $s_1$ is relatively underpopulated. Its interaction with cluster $2$, which contains a larger number of oscillators, makes its amplitude to rapidly decrease with $K$, as seen for $r_1=0.2$ in the upper set of curves of Fig.~\ref{fig3}. In contrast, the dependence on $K$ of the amplitude of cluster $2$ (lower set of curves) is more moderate. The rapid change with $K$ in the oscillation amplitude of cluster $1$ leads to its encounter with the continuation of the unstable solution $u$, causing the disappearance of $S_{12}$ by merging with $S_{u2}$. For large $r_1$, on the other hand, most oscillators are in cluster $1$. It is now the amplitude of cluster $2$ which varies rapidly with $K$ (see the lower curve for $r_1=0.8$ in Fig.~\ref{fig3}). As a consequence, the continuations of $s_2$ and $u$ encounter each other, and disappearance occurs by merging of $S_{12}$ and $S_{1u}$.

The distribution of oscillators between the two clusters also plays a role in the outcome of destabilization as $K$ grows. This is clearly seen from the results shown in Fig.~\ref{fig5}. In the low-$r_1$ regime, where most oscillators are in cluster $2$, destabilization at $K_U$ leads the whole ensemble to synchronize on solution $s_2$ (open dots in the figure). Conversely, for large $r_1$, most oscillators are in cluster $1$, and destabilization makes the system to collapse to solution $s_1$ (full dots). This general trend, however, is strongly modulated by the value of the excitation frequency $\Omega$. As $\Omega$ becomes larger, due to the increasing proximity of the unstable state $u$, the basin of attraction of $s_1$ decreases by comparison with that of $s_2$. As a result, the value of $r_1$ necessary to switch between the two regimes grows --namely, a larger and larger fraction of the ensemble must lie in cluster $1$ for its dynamics to prevail in the collective behavior of the system. 

\section{Conclusion} \label{sec4}

We have studied the collective dynamics of an ensemble of globally coupled, externally forced, identical Duffing oscillators, within their bistable regime. From the leading order in the multiple-scale approximation, we have shown the existence of stationary solutions where the ensemble is split into two clusters. Within each cluster all oscillators are synchronous and follow the same orbit, while the two clusters generally have different numbers of oscillators, amplitudes and phases. These two-cluster solutions exist as a consequence of the bistability of each individual oscillator. They are expected to disappear as the coupling between oscillators becomes strong enough and the whole ensemble collapses into a fully synchronized state. Since, for weak coupling, two-cluster oscillations exist for any possible splitting of the ensemble into two groups, this class of solutions represent a large multiplicity in the states accessible to the system.   

In a multiple-scale harmonic approximation, we have shown that there is a critical value $K_D$ of the coupling strength above which  two-cluster oscillations cease to exist as a solution to our problem. Depending on the fraction of oscillators in each cluster, this critical phenomenon is due to the merging with a different two-cluster state, which is a continuation of an unstable solution at $K=0$. This difference defines two well differentiated regimes, mediated by an abrupt jump on the threshold $K_D$.

When comparing these results with numerical integration of the equations of motion, we have found systematic discrepancies between the values of $K_D$ and the coupling strength $K_U$ at which the two clusters collapse into a single one. These discrepancies derive from the fact that, while in our multiple-scale approximation only the coordinates of the two clusters are involved, the numerical integration accesses the dynamics of all individual oscillators.  The selection of the final state into which the two clusters collapse upon destabilization mainly depends on a balance between the proportion of oscillators in each cluster, and the distance of the final state to the unstable solution. While the presence of a large fraction of the ensemble in one of the clusters favors the collapse into the closest stable state, the proximity of the unstable solution may cause the system to collapse into the farthest one.

The ensemble of coupled identical Duffing oscillators studied here provides an example of multistable self-organized clustering stemming from the individual dynamics of each oscillator. Besides their intrinsic interest for the field of nonlinear dynamics, our results may be relevant to the technology of micromechanical devices \cite{Lif}, where the Duffing oscillator is a paradigmatic model for the motion of pacemaking components. 

\bibliographystyle{ws-ijbc}
\bibliography{sosa-duf}
\end{document}